\documentclass[pra, aps, twocolumn, nofootinbib, tightenlines, nobibnotes, showpacs]{revtex4-1}
\usepackage{bigints}
\usepackage{amssymb}
\usepackage{graphicx}
\usepackage{amsmath}
\usepackage{dcolumn}
\usepackage{bm}
\usepackage{multirow}
\usepackage{hyperref}
\usepackage{physics}
\usepackage{bm}
\usepackage{comment}
\usepackage{xcolor}
\usepackage[caption=false]{subfig}
\usepackage{float}
\usepackage{siunitx}
\usepackage{dsfont}
\usepackage{appendix}
\usepackage{bbold}

\definecolor{lapislazuli}{rgb}{0.15, 0.38, 0.61}
\definecolor{YKblue}{rgb}{0.0, 0.18, 0.65}
\definecolor{carmine}{rgb}{0.81, 0.09, 0.13}
\definecolor{lavender}{rgb}{0.84, 0.79, 0.87}

\hypersetup{
	colorlinks=true,
	linkcolor=YKblue,
	linktoc=page,
	citecolor=YKblue,
	urlcolor=YKblue
}

\usepackage{tikz}
\usetikzlibrary{patterns,arrows,shapes,arrows,intersections,decorations}

\begin{document}
	
	
	\title{Experimental implementation of quantum gates with one and two qubits using Nuclear Magnetic Resonance}
	
	\date{September 2020}

	\author{Jos\'e L. Figueiredo}
	\email{jose.luis.figueiredo@tecnico.ulisboa.pt} 
	\affiliation{Instituto de Plasmas e Fus\~{a}o Nuclear, Instituto
	Superior T\'{e}cnico, Universidade de Lisboa, 1049-001 Lisboa, Portugal}

\begin{abstract}

Nuclear Magnetic Ressonance (NMR) is a widely used technique, with a long history of applications in chemestry, medicine, and material science. Twenty years ago, it emerged as a reliable source for quantum computing too, since the work of Cory et al. \cite{Cory}. One of its major advantage is the ease with which arbitrary unitary transformations can be implemented, together with its experimental simplicity, that can be traced back to very simple NMR routines, which were being extensively used long before. However, some disadvantages came along, mostly related to experimental effort in the initialisation and measure processes, and scalability. In this work, we have successfully probed some simple quantum gates (\textit{Pauli-Z}, \textit{Pauli-X} and \textit{Hadamard}) in one and two-qubit systems, realised in a NMR experiment. The work comprised a pseudo-pure state preparation, followed by the application of the gates, and a quantum tomography method, necessary to reconstruct the density matrix. The experiments were conducted with a chloroform sample, placed in a $300$ MHz Bruker Avance II spectrometer, equipped with a superconducting magnet with a $7$ T magnetic field. 
 	
\end{abstract}	


\maketitle
\section{Introduction and theoretical background}
Information processing technologies, like communication and computation, can be found among the most defining features of the last decades, and have been used in many areas of our daily lives. Computation, in particular, has gone through an extraordinary progress, and is still evolving at a very fast speed. However, this ongoing progress is rapidly reaching its maximum, by colliding with some fundamental limits: the quantum scale. This limit is approached as the electronic components decrease its size, and reach the atomic scale. Once at this scale, we are not able to rely on the classical laws of physics anymore, because of delocalization \cite{quantum1,quantum2}. Therefore, a wave-function formalism is necessary to describe matter and its interactions. One way to solve the problem is to use devices that explicitly operate in this quantum fashion. Researchers soon realised that such devices would be able to do much more: they could also use such quantum properties to perform tasks that are way beyond the capabilities of the previous classical computers \cite{2}. Although presenting new challenges \cite{3}, this provides a fundamental solution to a large set of obstacles that have undermined electronic experiments, and all sort of communication and computation systems. \par 
 The theoretical background of quantum information has been the focus of many theoreticians over the years, which lead to a rapid development of the field, becoming one of the hottest topics of research. Initially it appeared that actually building a quantum computer would be extremely difficult, but in the last few years there has been an explosion of interest in the use of techniques adapted from conventional liquid state NMR experiments to build small quantum computers.\par 
 An NMR quantum computer is, however, fundamentally different from other more conventional approaches. Traditional quantum computers are designed to be comprised of N two level quantum systems, with some coupling interactions between them (usually, with the first neighbours), and possible interactions with some exterior environment. On the other hand, NMR uses a large number of independent quantum computers, codified in each of the molecules in a given sample, which transform coherently in some temporal scale. Because molecules are largely isolated in liquid samples, the observable NMR spectrum is the sum of all the independent contributions, in a synchronous and parallel manner. Thus, such a machine performs a calculation using quantum parallelism at the molecular level and then amplifies its results to the macroscopic level via a form of classical parallelism. Such a quantum computer is many times called an ensemble quantum computer \cite{cory2}, because it uses a thermal ensemble of nearly independent systems.\par 
  NMR is the study of the direct transitions between the Zeeman levels of an atomic nucleus in a magnetic field, thus providing a very simple toy model of quantum behaviour \cite{b1,b2}. Many atomic nuclei possess an intrinsic angular momentum, called spin ($\mathbf{S}=\hbar\mathbf{I}$), and thus an intrinsic magnetic moment ($\bm{\mu}$), related by $\bm{\mu}=\gamma\hbar\mathbf{I}$, where $\gamma$ is the gyromagnetic ratio, and is a constant which depends on the nuclear species. In a suitable representation, $\mathbf{I}$ is given by the Pauli matrices divided by 2, i.e., $\mathbf{I}=\bm{\sigma}/2$, where $\bm{\sigma}=(\sigma_x,\sigma_y,\sigma_z)$ and 
  \begin{equation}
  	\sigma_x = \mqty(0 & 1 \\ 1 & 0), \quad  	\sigma_y = \mqty(0 & -i \\ i & 0), \quad  \sigma_z = \mqty(1 & 0 \\ 0 & -1).
  \end{equation}
If the nucleus is placed in a magnetic field, the spin will be quantised, with a small number of allowed orientations with respect to the field. For both conventional NMR and NMR quantum computing the most important nuclei are those with a spin-$1/2$: these have two spin states, which are separated by the Zeeman splitting
\begin{equation}
	\Delta E = \hbar \gamma B,
\end{equation}
in the presence of a magnetic field $B$. This separation is due to the fact that, under a magnetic field, a spin acquires an energy term (the so called Zeeman term) of the form
\begin{equation}
	\mathcal{H}_\text{Zeeman} = -\bm{\mu}\cdot\mathbf{B}. \label{zeeman}
\end{equation}
Transitions between the Zeeman levels can be induced by an oscillating magnetic field with a resonance frequency $\omega_0 = \Delta E/\hbar$ (the Larmor frequency). In the presence of a magnetic field $\mathbf{B}$, the rate of change in $\mu$ is governed by Bloch's equation
\begin{equation}
	\frac{d \bm{\mu}}{dt} = \gamma \bm{\mu}\times \mathbf{B}.
	\label{1}
\end{equation}
A solution of Eq.\eqref{1} depicts a precession movement, at a frequency $\omega = \omega_0$, of the magnetic moment around the direction of $\mathbf{B}$.

\subsection{The NMR experiment}
In a typical NMR experiment, nuclei are placed under a strong and static magnetic field 
\begin{equation}
	\bm{B}_0 = B_0 \bm{e}_z,
\end{equation}
and a weak perpendicular magnetic field (called the \textit{radio-frequency} magnetic field)
\begin{equation}
	\bm{B}_1 = B_1\cos(\omega_{rf}t + \phi) \bm{e}_x -  B_1\sin(\omega_{rf}t + \phi) \bm{e}_y,
\end{equation} 
such that $\abs{B_0}\gg \abs{B_1}$. This process occurs near resonance, when the oscillation frequency matches the Larmor frequency of the nuclei. In the spectrometer we read the transverse magnetization
	\begin{equation}
		M_{xy} = \text{Tr}\big[\rho\sum_i I_+^i \big]/Z, \label{ob}
	\end{equation}
where $I_+^i = I_x^i +iI_y^i$ is the raising operator, $\rho=\ket{\psi}\bra{\psi}$ is the density matrix, $Z=\Tr \rho$, and the summation is performed over the different $n$ spins in the same molecule. The typical shape of the magnetization is a sum of peaks, in the frequency domain (see Fig.\ref{spectra}). Eq.\eqref{ob} was first derived by Cory, who realised that the $n$-particle quantum mechanics could be stored in a much reduced wave function, with the dimensionality of the $n$ spins in each molecule \cite{Cory}. \par
Furthermore, in order to perform large quantum computations, it is necessary to ensure that the decoherence time is long in comparison with the time required to implement a quantum logic gate, so that it is possible to implement many logic gates before any significant decoherence occurs \cite{times}. In liquid samples, two characteristic time scales are important in quantum computing:
\begin{itemize}
		\item $T_1$, spin-lattice relaxation time, measures how fast the longitudinal magnetization $M_{z}$ goes back to the equilibrium configuration. It originates from couplings between the spins and the lattice, that is, excitation modes which can carry away energy quanta on the scale of the Larmor frequency. In general, a phenomenological equation that governs the $z-$component of the magnetization is given by
		\begin{equation}
			M_z(t)=M_{z,eq}-\big[M_{z,eq}-M_z(0)\big]e^{-t/T_1}\ ,
		\end{equation}
		i.e, the magnetization recovers to 63\% of its equilibrium value after one time constant $T_1$.
		\item  $T_2$, spin-spin relaxation time, measures how fast the transverse magnetization $M_{xy}$ disappears, being created by the pulses. It originates from spin-spin couplings which are imperfectly averaged away, or unaccounted for in the Hamiltonian. For example, in molecules in liquid solution, spins on one molecule may have a long range, weak interaction with spins on another molecule. Fluctuating magnetic fields, caused by spatial anisotropy of the chemical shift or unstable laboratory fields, also contribute to $T_2$. The corresponding phenomenological equation is
		\begin{equation}
		M_{xy}(t)=M_{xy}(0)e^{-t/T_2} \ ,
		\end{equation}
		i.e., the transverse magnetization vector drops to 37\% of its original magnitude after one time constant $T_2$.
\end{itemize}
It is, therefore, important to make sure that the duration of the experiment does not exceed those two characteristic time scales (which is easily the case, once those times, in low viscosity liquids are large, of the order of the hundreds of milliseconds).

\subsection{The system Hamiltonian}
\subsubsection{One-qubit system}
\begin{figure}[H]
		\centering
		\includegraphics[scale=0.42]{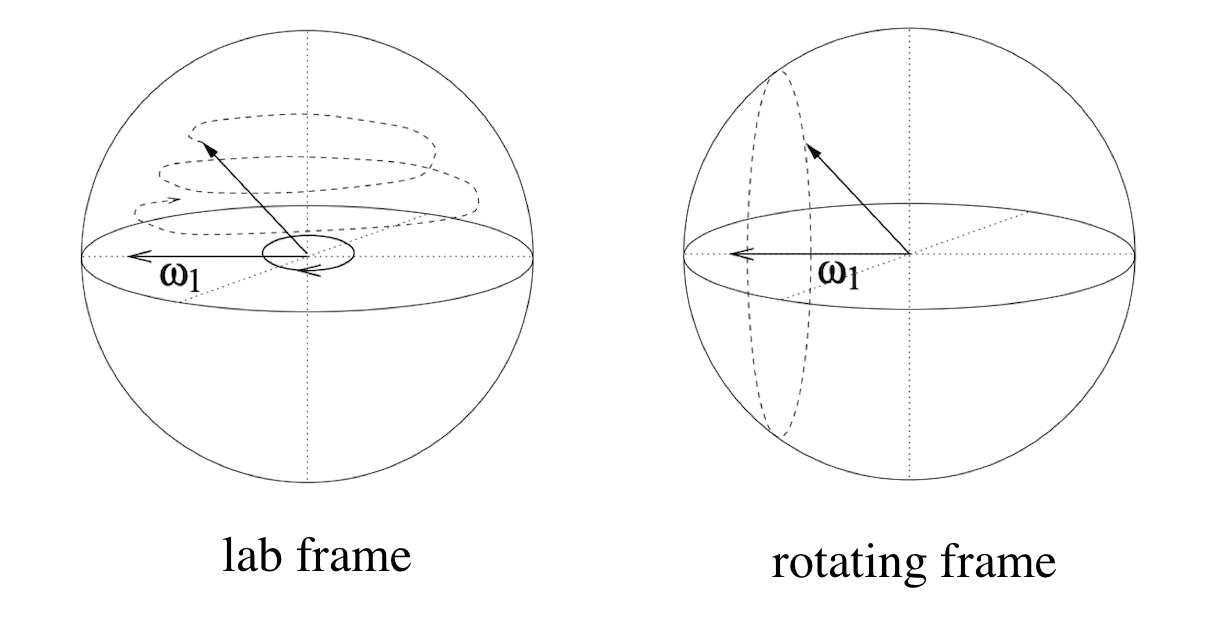}
		\caption{Different motion of the spin in two distinct frames - in the left panel, the spin is precessing in the laboratory frame, whereas in the right panel, the spin is simply rotating, in the rotation frame. The energy is time-dependent and time-independent, respectively, in each frame.}
		\label{rotFrame}
\end{figure}

One-qubit systems are realised in liquid samples whose molecules have only one spin. Therefore, no interaction between spins is considered. This system is the simplest quantum system of all, comprised of two states $\ket{0}$ and $\ket{1}$, denoting the eigenstates of the $I_z$ operator, such that 
\begin{align}
	I_z \ket{0} &= +\frac{1}{2}\ket{0},\\
	I_z \ket{1} &= -\frac{1}{2}\ket{1},
\end{align}
are the down and up state, respectively. A general vector in this space is 
\begin{equation}
	\ket{\phi} = \alpha \ket{0} + \beta\ket{1},
\end{equation}
where $\alpha$ and $\beta$ are complex coefficients, which verify the condition $\abs{\alpha}^2+\abs{\beta}^2=1$, or equivalently, $\bra{\phi}\ket{\phi}=1$. For the experimental conditions, the total Hamiltonian is given by
\begin{align}
	\mathcal{H} =&-\hbar \omega_0 I_z - \hbar \omega_1 \big[\cos(\omega_{rf}t + \phi)I_x \nonumber \\
	& - \sin(\omega_{rf}t + \phi)I_y\big], \label{th1s}
\end{align}
which comprises two Zeeman terms (see Eq.\eqref{zeeman}), such that $\omega_0 = \gamma B_0$ and $\omega_1 = \gamma B_1$ are the two Larmor frequencies of each magnetic field. The Schr\"{o}dinger equation,
\begin{equation}
i\hbar\partial_t \ket{\psi} = \mathcal{H}\ket{\psi},
\end{equation}
is hard to solve because $\mathcal{H}$ is time dependent. By performing the rotation (see Fig.\ref{rotFrame})
\begin{equation}
\ket{\psi}_{\text{rot}} = e^{-i\omega_{rf}t \  I_z}\ket{\psi},
\end{equation}
the new Schr\"{o}dinger's equation becomes 
\begin{equation}
i\hbar \partial_t \ket{\psi}_{\text{rot}}  = \mathcal{H}_{\text{rot}} \ket{\psi}_{\text{rot}},
\end{equation}
where $\mathcal{H}_{\text{rot}}$ is a time-independent Hamiltonian, given by
\begin{align}
	\mathcal{H}_{\text{rot}}[\omega_0,\omega_1,\phi] = &-\hbar (\omega_0-\omega_{rf})I_z  \nonumber \\
	&-\hbar\omega_1(\cos\phi I_x - \sin\phi I_y).
\end{align}
If we choose $\omega_{rf} = \omega_{0}$ (resonance) the $B_0-$Zeeman term vanishes, and the effective Hamiltonian reduces to a simpler expression,
\begin{equation}
	\mathcal{H}_{\text{rot}}[\omega_1,\phi] = -\hbar\omega_1(\cos\phi I_x - \sin\phi I_y).
\end{equation}
Therefore, through the application of radio-frequency pulses of a given width ($\omega_1\Delta t$) and phase ($\phi$), any quantum logic gate can be implemented, and measured in the rotating frame, 
\begin{equation}
 \ket{\psi(t)}_{\text{rot}}= e^{-i	t\mathcal{H}_{\text{rot}}[\omega_1,\phi]/\hbar} \ket{\psi(0)}_{\text{rot}}.\label{3}
\end{equation}
	
\subsubsection{Two-qubit system}
For two-qubit systems the approach is similar, although the Hilbert space is composed by four states $\{\ket{00},\ket{01},\ket{10},\ket{11}\}$,, written in the computational basis. Note that 
\begin{equation} 
 \ket{\alpha \beta} = \ket{\alpha}\otimes\ket{\beta},
  \end{equation} 
  where $\ket{\alpha}$ and $\ket{\beta}$ refer to each individual spin states, respectively. The total Hamiltonian in the laboratory frame will include time-dependent terms, which can't be treated exactly, but rather perturbatively, which is not ideal for our purpose. Therefore, in order to find a solution as the one depicted in Eq.\eqref{3}, we will move to the so called double-rotating frame. We will denote the two different spins with a superscript $i=\{1,2\}$, where $1$ stands for the hydrogen spin, and $2$ stands for the carbon-13 spin (e.g., the spin operator in the $z-$direction for the hydrogen is denoted by $I_z^1$). Note that every operator must be extended to the $2^2=4$-dimensional Hilbert space, by performing a tensorial product with the identity operator for the remaining spin (e.g., the effective spin operator in the $z-$direction for the hydrogen is, in fact, $I_z^1\otimes\mathbb{1}^2$). In the spectrometer, two different channels are settled, one for each spin. This means that, instead of just one, there will be two oscillating magnetic fields $\mathbf{B}_1^i$, one for each spin species, whose components are $\mathbf{B}_1^i = B_1^i\cos(\omega_{rf}^it + \phi^i) \bm{e}_x -  B_1^i\sin(\omega_{rf}^it + \phi^i) \bm{e}_y$. The $\mathbf{B}_0$-field is the same for each channel. The total Hamiltonian is now
\begin{align}
	\mathcal{H} =&-\hbar \omega_0^1 I_z^1\otimes\mathbb{1}^2 -\hbar \omega_0^2 \mathbb{1}^1\otimes I_z^2 \nonumber \\
	&- \hbar \omega_1^1 \Big[\cos(\omega_{rf}^1t + \phi^1)I_x^1\otimes \mathbb{1}^2 \nonumber \\
	& - \sin(\omega_{rf}^1t + \phi^1)I_y^1\otimes \mathbb{1}^2\Big] \nonumber \\
	&- \hbar \omega_1^2 \Big[\cos(\omega_{rf}^2t + \phi^2) \mathbb{1}^1\otimes I_x^2 \nonumber \\
	& - \sin(\omega_{rf}^1t + \phi^2)\mathbb{1}^1\otimes I_y^2\Big] \nonumber\\
	& +2\pi\hbar J\big(I_z^1\otimes I_z^2\big)\label{th2s}, 
\end{align}
where the variables $\omega_0^i$ and $\omega_1^i$ are defined as before, for each spin channel. The first terms on the RHS of Eq.\eqref{th2s} are the natural generalisation of the previous case in Eq.\eqref{th1s}. The difference is present in the last term, the $J-$coupling term, which mediates the interaction between the two spins in the same molecule. With the same reasoning as before, the transformation to the double-rotating frame is
\begin{equation}
\ket{\psi}_{\text{rot}} = e^{-i\omega_{rf}^1t \  I_z^1\otimes\mathbb{1}^2}e^{-i\omega_{rf}^2t \  \mathbb{1}^1\otimes I_z^2}\ket{\psi}.
\end{equation}
Choosing $\omega_{rf}^1 = \omega_0^1$ and $\omega_{rf}^2 = \omega_0^2$ allows to remove the $\mathbf{B}_0-$Zeeman term, and the final Hamiltonian reads
\begin{align}
	\mathcal{H}_{\text{rot}}&[\omega_1^1,\phi^1,\omega_1^2,\phi^2] = - \hbar\omega_1^1\big(\cos\phi^1  \ I_x^1\otimes\mathbb{1}^2  \nonumber\\
	&- \sin\phi^1 \ I_y^1\otimes\mathbb{1}^2\big)-  \hbar\omega_1^2\big(\cos\phi^2 \  \mathbb{1}^1\otimes I_x^2\nonumber \\
	& - \sin\phi^2 \  \mathbb{1}^1\otimes I_y^2 \big) + \hbar  2\pi J \ ( I_z^1\otimes  I_z^2). \label{2spinH}
\end{align}
Then, any unitary transformations can be tuned with the parameters $\{t\omega_1^1,\phi^1,t\omega_1^2,\phi^2\}$, 
\begin{equation}
 \ket{\psi(t)}_{\text{rot}}= e^{-i	t\mathcal{H}_{\text{rot}}[\omega_1^1,\phi^1,\omega_1^2,\phi^2]/\hbar} \ket{\psi(0)}_{\text{rot}},
\label{uni2}
 \end{equation}
 as in the previous case.
\subsection{Density matrix and pseudo-pure states}\label{densitymatrix}
When an unitary transformation, such as the one in Eq.\eqref{3}, is applied on $\ket{\psi}$, i.e., $\ket{\psi} \rightarrow U\ket{\psi}$, its action on the density matrix is, accordingly, $\rho \rightarrow U \rho U^\dagger$. Furthermore, in NMR, it is much more convenient to work with the density matrix representation, rather than the state representation, because the NMR observable is easily calculated with $\rho$ (see Eq.\eqref{ob}). It is easy to show that both representations are equivalent. Consider a general state in the Hilbert space of a 4 state system (e.g. two-qubit system)
	\begin{equation*}
	\ket{\psi} = c_1 \ket{1} + c_2\ket{2}+c_3 \ket{3}+c_4 \ket{4} .
	\end{equation*}
	The density matrix can be reconstructed from the wave-function parameters as 
\begin{equation}
	 \rho = 
	\left( \begin{array}{cccc} 
\abs{c_1}^2 & c_1c_2^\ast & c_1c_3^\ast & c_1c_4^\ast\\
c_2c_1^\ast & \abs{c_2}^2 &  c_2c_3^\ast &  c_2c_4^\ast \\
c_3c_1^\ast &  c_3c_2^\ast  &   \abs{c_3}^2 & c_3c_4^\ast  \\
c_4c_1^\ast & c_4c_2^\ast & c_4c_3^\ast & \abs{c_4}^2 \\
	 \end{array} \right) ,
\end{equation}
where, in our notation, the operator $\ket{1}\bra{1}$ corresponds to the first row and first column, the operator $\ket{1}\bra{2}$ corresponds to the first row and second column, and so on (the states grow in number as we move to the right in the row and to the bottom in the column). We can also reconstruct the state from the density matrix. To do that, we need to find the eight real numbers that correspond to the absolute value $\abs{c_i}$ and phase $\theta_i$ of each $c_i$ element, because $\rho=\rho^\dag$. The absolute values are directly extract from the diagonal elements
\begin{equation}
	\abs{c_i} = \sqrt{\rho_{ii}} .
\end{equation}
For the phases, it is enough to obtain the phase diferences $\theta_i - \theta_j$ for some choosen $j$, given that the state $\ket{\psi}$ is defined up to an overall phase. We can choose $j=4$, i.e., multiply $\ket{\psi}$ by $e^{-i\theta_4}$, and in that case 
\begin{align*}
	\theta_1 &= Arg(\rho_{14}),\\
	\theta_2 &= Arg(\rho_{24}),\\
	\theta_3 &= Arg(\rho_{34}),\\
	\theta_4 &= 0,
\end{align*}
where $Arg(z)$ returns the phase of $z$. This procedure is trivially generalised for any dimension of the Hilbert space.\par
Moreover, in thermal equilibrium, the density matrix is given by Boltzmann's distribution
\begin{equation}
\rho= e^{-\mathcal{H}/k_BT}\approx\mathbb{1} -\mathcal{H}/k_BT, \label{4}
\end{equation}
where the high temperature approximation is valid for the laboratory conditions.\par
A conventional quantum computer starts its calculations from the pure ground state, which for a two qubit device can be written as $\ket{00}$. This state is not normally accessible to NMR experiments, but it suffices to start from a so called pseudo-pure state, that behaves in a similar fashion in the NMR experiment. It is written generically as 
\begin{equation}
	\rho= \xi_1\mathbb{1} + \xi_2\ket{n}\bra{n}, \label{2}
\end{equation}
where $\mathbb{1}$ is the $2^N\times 2^N$ identity matrix, $N$ is the dimensionality of the Hilbert space, and $\xi_{1,2}$ are temperature-dependent coefficitens, and $\ket{n}$ is a state of the Hilbert space (usually, the ground state). The second term of Eq.\eqref{2} is called the deviation density matrix, and is responsible for the detectable NMR signal \cite{pseudo}. Therefore, NMR is capable of detecting signals coming from nuclei in the same quantum state over a maximum entropy background. Once the matrix representation of the $I^+$ operator is traceless, it is easy to see that, after the unitary transformations of Eq.\eqref{3} or \eqref{uni2}, the measured transverse magnetization will only be affected by the second term on the RHS of Eq.\eqref{2}, and so it reads
\begin{equation}
	M_{xy}(t) = \xi_2 \Tr\Big[\sum_i U(t)\ket{n}\bra{n} U(t)^\dagger I_+^ i\Big]/Z,
\end{equation}
which is equivalent to applying an unitary transformation on a pure state. From now on, $\rho$ will be used to represent the deviation matrix, with zero trace.
\section{Quantum state tomography}
\begin{figure}[H]
	\centering
	\includegraphics[scale=0.3]{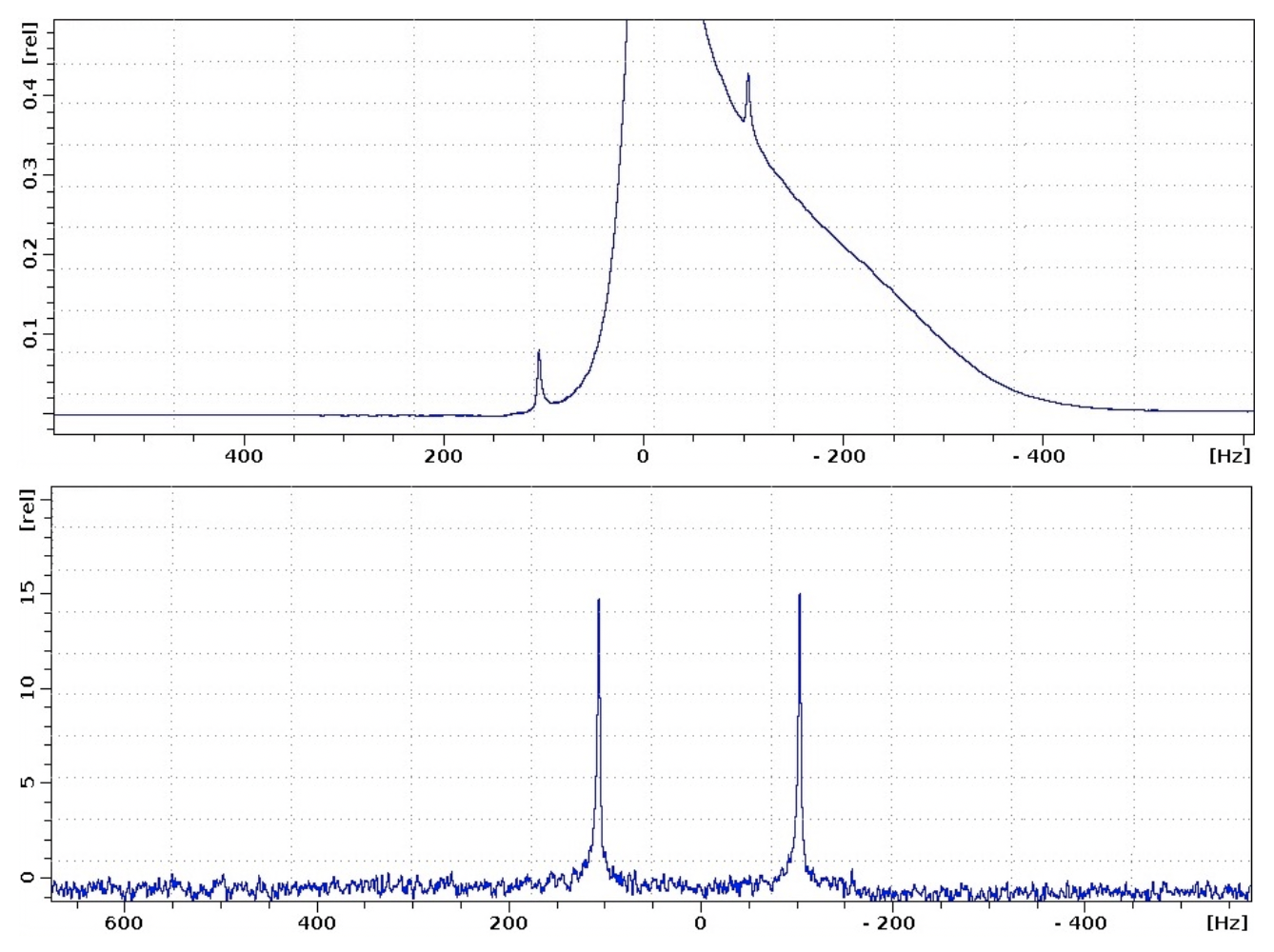}
	\caption{Experimental (Fourier transformed) NMR spectra, after a $\pi/2-$pulse applied in the $x-$direction, in both channels. The horizontal axis is frequency (in Hz), the vertical axis is in arbitrary units, and the top and bottom panels are, respectively, the hydrogen and carbon-13 channels. The first panel contains data from both the one-qubit and two-qubit experiments, as they are impossible to isolate in the raw data. The one-qubit magnetization contains one single peak (the one centred around $\omega =0$), and the two-qubit magnetization contains two peaks (in the top panel), centred in $\omega=\pm\pi J$, separated by $2\pi J$; and two other peaks (in the bottom panel),  centred in $\omega=\pm\pi J$, separated by the same amount. The frequency increases from right to left, and is shifted by the Larmor frequencies of each channel $i$ ($\omega =  \omega' + \omega_0^i$, where $\omega'$ is the horizontal axis).}
	\label{spectra}
\end{figure}

Tomography refers to a set of operations we need to perform, in order to fully reconstruct $\rho$. The Fourier transform of the observable $M_{xy}(\omega)$ is, in the ideal case,
\begin{align}
	 M_{xy}(\omega)&\sim \int \dd t \ e^{i\omega t} \ Tr[ \ \rho(t)I^{+} ] \nonumber \\
	 &= \sum_{i} \Gamma_i(\rho) \ \delta(\omega-\Delta E_i/\hbar), \label{cete} 
\end{align}
where $i$ runs over all possible single spin transitions, and $\{\Gamma_i\}$ are a set of coefficients that are related to some $\rho_{ij}$ elements, which are accessible after integrating $M_{xy}(\omega)$ around each delta peak (see Fig.\ref{spectra}). To access the remaining elements, a set of unitary transformations $\{\Pi_i\}$ must be applied to the system, which will change Eq. \eqref{cete} as $\rho\rightarrow \Pi_i\rho\Pi_i^\dag$. The new $\Gamma$ functions featuring in Eq.\eqref{cete}, corresponding to the amplitude of each frequency-peak, are a linear combination of different density matrix elements, and a finite number of operations permits to reconstruct $\rho$. Depending on the number of qubits, this set of transformations may become very large; however, for the present case, the number of operations is still treatable. 

\subsection{Tomography of a one-qubit system}
For a system composed by one qubit only, there is only one single qubit transition, centred around $\omega=\omega_0^1\approx 2\pi\times 300$ MHz (the hydrogen frequency), and so one $\Gamma$ coefficient, 
\begin{equation}
\Gamma_1= \rho_{21} = \rho_{12}^\ast. \label{c1}
\end{equation}
The element of Eq.\eqref{c1} can be readly obtained by integrating the spectrum, separately, for the real and imaginary parts. In this case, a $\pi/2$ rotation around the $x-$direction, denoted by $\Pi_1$, is enough to obtain the other elements. This is realised by the transformation of Eq.\eqref{3} if one chooses the control parameters to be $\Delta t\omega_1 = \pi/2$ and $\phi=0$, 
\begin{equation}
	\Pi_1 = e^{-i\frac{\pi/2}{\omega_1}\mathcal{H}_{\text{rot}}[\omega_1,0]/\hbar}.
\end{equation}
After the application of the pulse, the amplitude of the single peak becomes
\begin{equation}
\Gamma_1= i(\rho_{11}-\rho_{22}) + \text{Re}(\rho_{12}) \label{c2}.
\end{equation}
Eq.\eqref{c1} and \eqref{c2}, together with 
\begin{equation}
	\text{Tr}[\rho]=0,
	\label{c3}
\end{equation}
valid for a pseudo-pure state, are enough to fully reconstruct the deviation density matrix $\rho$.
\begin{figure}[H]
	\centering
	\includegraphics[scale=0.4]{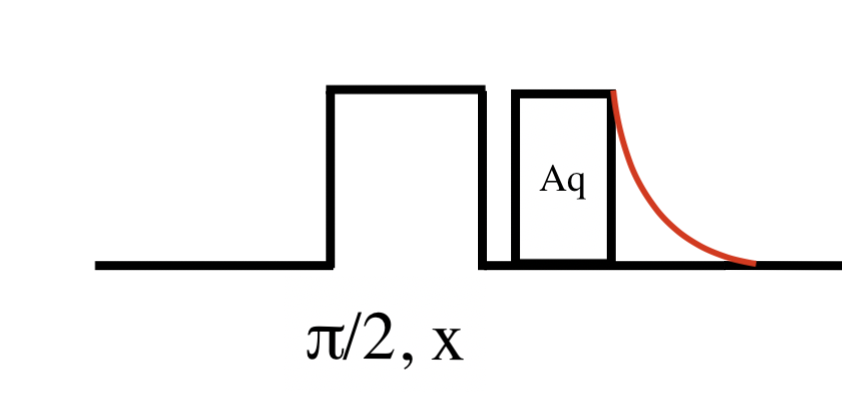}
	\caption{Tomography routine for the case of a single spin. It comprises one single pulse of $\pi/2$ in the $x-$direction ($\Pi_1$).}
\end{figure}

\subsection{Tomography of a two-qubit system}
 For two qubits, there are four single spin transitions, whose amplitudes are given by
\begin{align}
\Gamma_1= \rho_{13} = \rho_{31}^\ast, \label{gg1}\\
\Gamma_2= \rho_{24} = \rho_{42}^\ast,\\
\Gamma_3= \rho_{12} = \rho_{21}^\ast,\\
\Gamma_4= \rho_{34} = \rho_{43}^\ast.
\end{align}
This peaks are centred, respectively, around $\omega_0^1 + J$, $\omega_0^1 - J$, $\omega_0^2 + J$ and $\omega_0^2 - J$, where $\omega_0^1 \approx  2\pi\times 300$ MHz, and $\omega_0^2 \approx 2\pi\times 75.5$ MHz, are the hydrogen and carbon-13 Larmor frequencies associated with the $\mathbf{B}_0$ magnetic field. \par
\begin{figure}[H]
	\centering
	\includegraphics[scale=0.21]{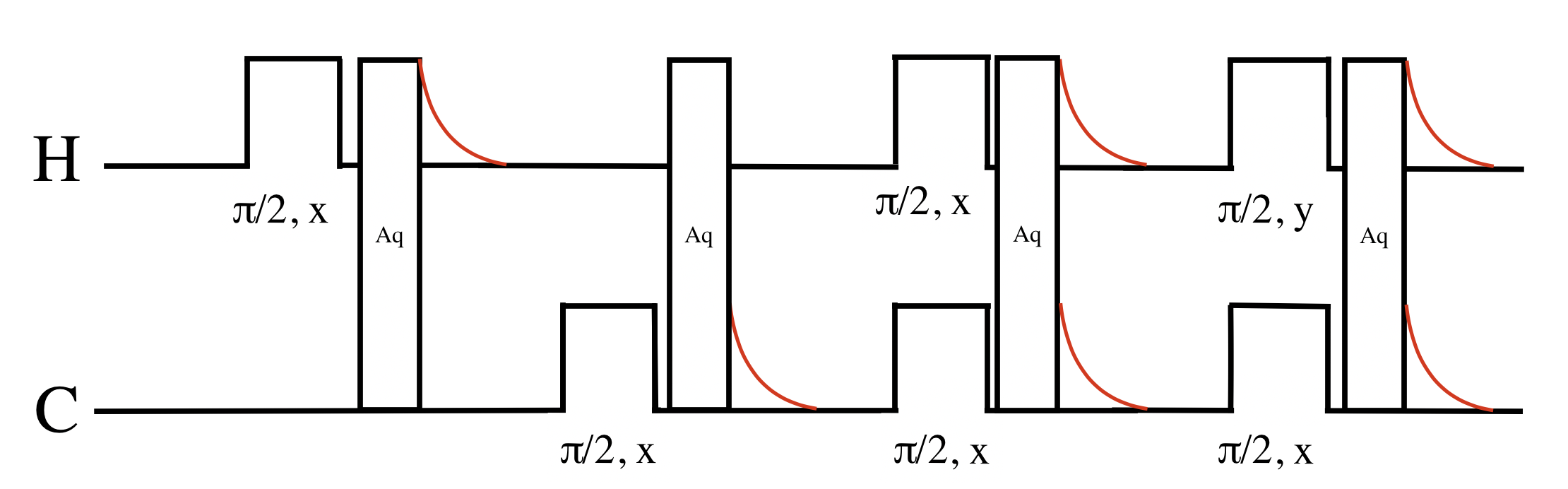} 
	\caption{Tomography routine for the two spins. They consist of two isolated pulses of $\pi/2$ in the $x-$direction in each of the channels, two pulses of $\pi/2$ in the $x-$direction simultaneously in the two chanels, and a pulse of $\pi/2$ in the $y-$direction in the hydrogen channel and a pulse of $\pi/2$ in the $x-$direction in the carbon channel, simultaneously.}
	\label{tom2FIG}
\end{figure}

In this case, the tomography procedure is way more laborious. There isn't only one set of pulses that allows the reconstruction of $\rho$, and we should choose the method that best suits our laboratory conditions. Now, we denote the set of operations by $\{\Omega_i\}$, represented in Fig.\ref{tom2FIG}, with time durations $\Delta \tau$. Their expressions, as well as the relevant $\Gamma$ function(s) of each step, are given below.
\begin{itemize}
	\item $\pi/2$ in the $x-$direction in the hydrogen channel ($\Delta \tau_1 \ \omega_1^1 = \pi/2$)
\end{itemize}
\begin{equation}
	\Omega_1=e^{-i\Delta \tau_1\mathcal{H}_{\text{rot}}[\omega_1^1,0,0,0]/\hbar },
\end{equation}
\begin{equation}
	\Gamma_2 = 2\rho_{11} + 4\rho_{22} + 2\rho_{33}. 
\end{equation}

\begin{itemize}
	\item $\pi/2$ in the $x-$direction in the carbon channel ($\Delta \tau_2 \ \omega_1^2 = \pi/2$)
\end{itemize}
	\begin{equation}
	\Omega_2=e^{-i\Delta \tau_2\mathcal{H}_{\text{rot}}[0,0,\omega_1^2,0]/\hbar },
\end{equation} 
\begin{align}
	\Gamma_3 &= 2\rho_{11} - 2\rho_{22},\\
	\Gamma_4 &= 2\rho_{11} + 2\rho_{22} + 4\rho_{33}.
\end{align}

\begin{itemize}
	\item $\pi/2$ in the $x-$direction in the hydrogen and carbon channel simultaneously ($\Delta \tau_3 \ \omega_1^1 = \pi/2$, $\Delta \tau_4 \ \omega_1^2 = \pi/2$ )
\end{itemize}
\begin{equation}
	\Omega_3=e^{-i\Delta \tau_4\mathcal{H}_{\text{rot}}[0,0,\omega_1^2,0]/\hbar }e^{-i\Delta \tau_3\mathcal{H}_{\text{rot}}[\omega_1^1,0,0,0]/\hbar },
\end{equation}
\begin{align}
	\text{Re}(\Gamma_1-\Gamma_2) &= 4\big[\text{Im}(\rho_{14})-\text{Im}(\rho_{23})\big],\\
	\text{Re}(\Gamma_3-\Gamma_4) &= 4\big[\text{Im}(\rho_{14})+\text{Im}(\rho_{23})\big],
\end{align}

\begin{itemize}
	\item $\pi/2$ in the $y-$direction in the hydrogen channel and $\pi/2$ in the $x-$direction in the carbon channel simultaneously ($\Delta \tau_5 \ \omega_1^1 = \pi/2$, $\Delta \tau_6 \ \omega_1^2 = \pi/2$ )
\end{itemize}
\begin{equation}
	\Omega_4=e^{-i\Delta \tau_6\mathcal{H}_{\text{rot}}[0,0,\omega_1^2,0]/\hbar }e^{-i\Delta \tau_5\mathcal{H}_{\text{rot}}[\omega_1^1,\pi,0,0]/\hbar },
\end{equation}
\begin{align}
	\text{Re}(\Gamma_1-\Gamma_2) &= 4\big[\text{Re}(\rho_{14})-\text{Re}(\rho_{23})\big],\\
	\text{Re}(\Gamma_4-\Gamma_3) &= 4\big[\text{Re}(\rho_{14})+\text{Re}(\rho_{23})\big],\label{gg2}
\end{align}
Note that, during the application of pulses, we can discard the $J-$coupling of $\mathcal{H}_\text{rot}$ (an argument for this is given shortly). Using the fact that $\rho=\rho^\dag$, it is easy to see that Eq.\eqref{gg1}-\eqref{gg2} fully determine $\rho$. Note that, after each step, not all the spectral information is necessary, and using more peaks would become redundant. That is the reason why, after each step, only somme $\Gamma$ functions are used. As it was already mentioned, there are several ways to set such a procedure. A more detailed description of quantum state tomography can be found elsewhere \cite{tom1,tom2,tom3}.
\section{Experimental method and results}
\begin{figure}[H]
		\centering
		\includegraphics[scale=0.28]{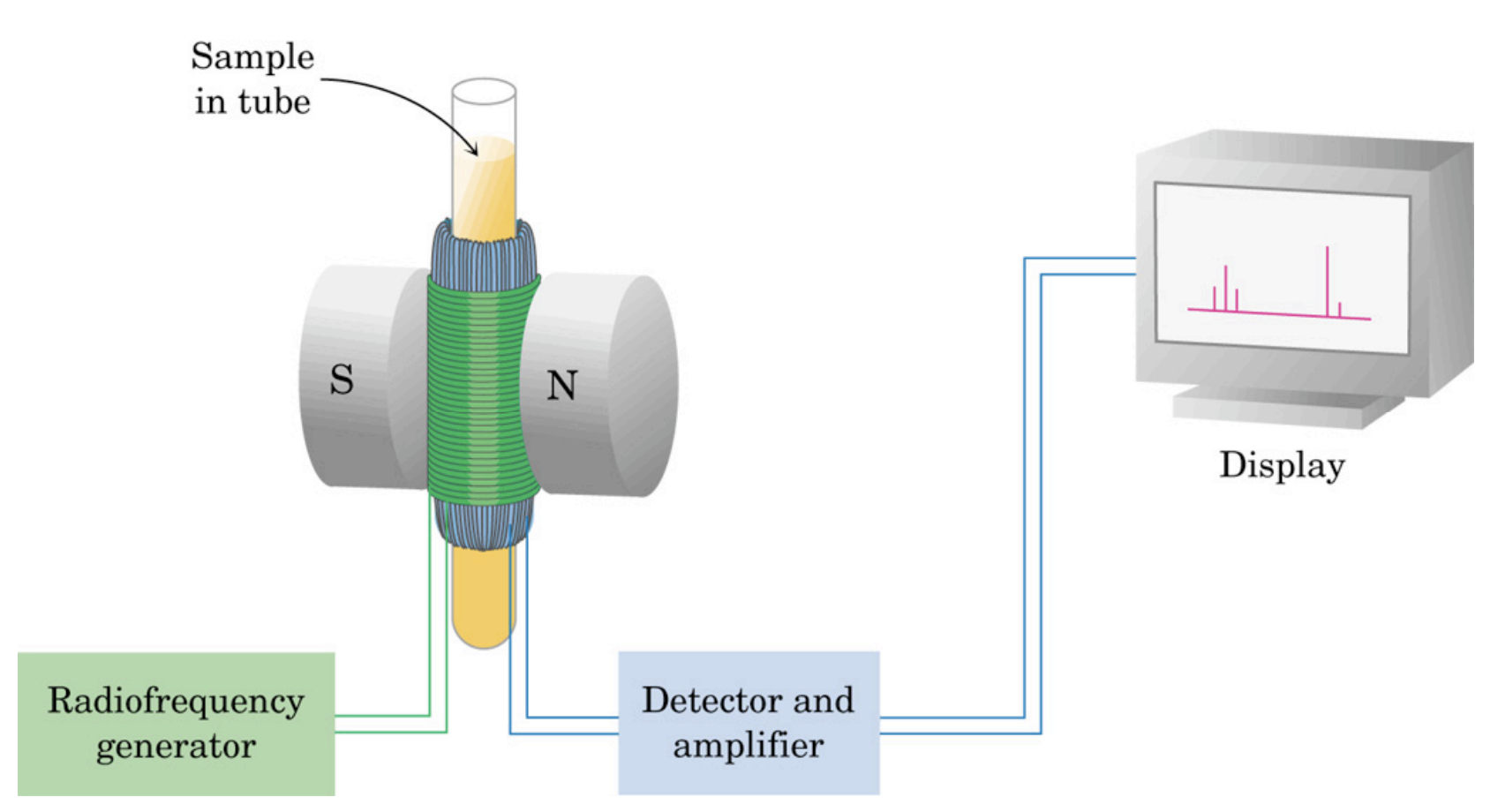}
		\caption{Experimental set-up, consisting of the sample inside the spectrometer, the superconducting magnet, and the display.} 
		\label{atuamae}
	\end{figure}

Two distinct cases were analysed: a one-qubit system was probed using the carbon-12 chloroform molecules, and a two-qubit system using the remaining carbon-13 chloroform molecules. 
\begin{figure}[H]
		\centering
		\includegraphics[scale=0.25]{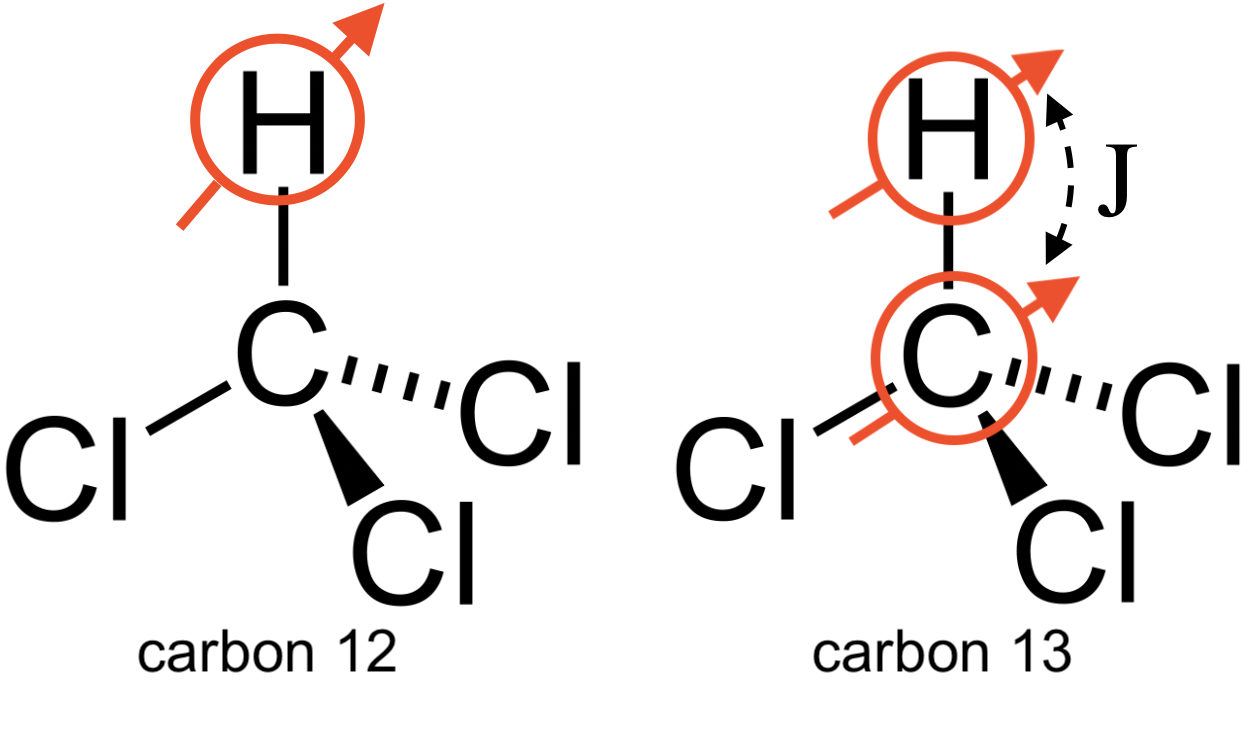}
		\caption{Two distinct molecules present in a chloroform sample. The chlor nucleus has no spin. The hydrogen (proton) nucleus is a spin$-1/2$ nucleus, whereas the carbon nucleus has a spin$-0$ isotope (carbon-12) and a spin$-1/2$ isotope (carbon-13), which coexist in the sample. Their relative abundances are $99\%$ and $1\%$, respectively.}
	\end{figure}	
The experimental set up is depicted in Fig.\ref{atuamae}. It consists of a 300 MHz Bruker Avance II spectrometer, composed by a superconducting magnet, a radio-frequency generator and receiver section. The superconducting magnet produces the strong magnetic field $\mathbf{B}_0$ in the $z-$direction, inside of which the sample is placed. A superconductor is used because it is capable of supporting large electric currents without any external source of power. Once charged with current, a superconducting magnet runs almost indefinitely, providing an extremely stable magnetic field with no outside interference. The magnet is also provided with two sets of additional coils, called \textit{shims}, for adjusting the homogeneity of the magnetic field. One set of coils, called the superconducting shims, is wound from superconducting material and immersed in the liquid-He bath. The radio-frequency generator (also called transmitter) section, is the part of the spectrometer responsible for the producing of the radio-frequency (r.f.) magnetic field. It includes the r.f. synthesizer that produces an oscillating electrical signal with a very well-defined frequency; a pulse gate, which is simply a fast switch that is opened at defined moments in order to allow the r.f. reference wave to pass through; a r.f. amplifier, which scales up the gated waveform so as to produce a large-amplitude r.f. pulse for transmission to the probe; and a probe, with which the sample is introduced inside the spectrometer. The receiver section includes preamplifier, which scales up the tiny signal to a more convenient voltage level; and a quadrature receiver, composed by analogue-to-digital converters (ADC), which converts the signal, to be observed in a computer. The results are analysed with a software called \textit{TopSpin}, also delivered by Bruker, which contains several features for treating the data, such as performing fast Fourier transform, coding several pulse programs, and graphically display the results. After collecting the data, some routines are constructed, in \textit{Mathematica}, to perform simple operations, such as solving linear systems, separate two different spectra out of the same data, and integration. Those are necessary to obtain the reconstructed matrix, $\rho^\text{exp}$.
\subsection{One-qubit system}
For the first case, the equilibrium density matrix of Eq.\eqref{4} can be approximated by,
\begin{align}
\rho_0  &\approx  \mqty(1 & 0 \\ 0 & 1)+ \frac{1}{k_BT}\mqty(\hbar\omega_0 & 0 \\ 0 & -\hbar\omega_0 ), \nonumber \\
&=  \Big(1-\frac{\hbar\omega_0}{k_BT}\Big) \ \mathbb{1}+ \frac{2\hbar\omega_0}{k_BT}\ket{0}\bra{0}. \label{rho}
\end{align}
Eq.\eqref{rho} is readly a pseudo-pure state, according to definition of Eq.\eqref{2}. \par 
 Some basic quantum logic gates were applied, realised by simple sequence pulses. We denote them by $S_1 = \sigma_x$ (the \textit{Pauli-X} gate), $S_2 = \sigma_z$ (the \textit{Pauli-Z} gate) and $S_3 = (\sigma_x + \sigma_z)/\sqrt{2}$ (the \textit{Hadamard} gate). Each of them is given by 
\begin{align}
	S_1 &= e^{-i\Delta t_1\mathcal{H}_{\text{rot}}[\omega_1,0]/\hbar}, \label{ff1} \\
	S_2 &= e^{-i\Delta t_1\mathcal{H}_{\text{rot}}[\omega_1,\pi/2]/\hbar} \  \ e^{-i\Delta t_1\mathcal{H}_{\text{rot}}[\omega_1,0]/\hbar}, \label{ff2} \\
	S_3 &= e^{-i\Delta t_1\mathcal{H}_{\text{rot}}[\omega_1,0]/\hbar} \ e^{-i\Delta t_2\mathcal{H}_{\text{rot}}[\omega_1,\pi/2]/\hbar}, \label{ff3}
\end{align}
and the conditions for $\Delta t_1$ and $\Delta t_2$ read
\begin{align}
	\Delta t_1 \omega_1^1 = \pi,\\
	\Delta t_2 \omega_1^1 = \pi/2.
\end{align}
Each choice of $\Delta t$ and $\phi_1$ corresponds to a r.f. pulse. Eq.\eqref{ff1}, for example, physically represents a $\pi-$pulse in the $x-$direction. A sequence of pulses can then be described by a matrix operator $S$, which transforms the density matrix as $\rho \rightarrow S\rho S^\dagger$. Furthermore, after the application of each sequence, the deviation (density) matrix is read with the tomography procedure described in the last section. As mentioned, a single frequency spectrum is not enough to reconstruct $\rho$, and a pulse sequence ($\Pi_1$) is necessary. Therefore, the tomography routine must receive two spectra, the one given by $S_i\rho S_i^\dag$, and the additional information contained in the spectrum of $\Pi_1S_i\rho S_i^\dag \Pi_1^\dag$. Then, a routine in \textit{Mathematica} was constructed, that receives this spectral information, solves the system of equations \eqref{c1}-\eqref{c3}, and returns the matrix $\rho$. The program is very simple, and we shall not describe it here, but, instead, denote it by $\mathcal{T}[..]$, where the arguments are the spectral information that is necessary. In the case of a single spin, as we have seen, the arguments are $\mathcal{T}[S_i\rho S_i^\dag,\Pi_1S_i\rho S_i^\dag \Pi_1^\dag]$, where $S_i$ is one of the quantum logic gates that were applied. \par 
To estimate the error ($\delta$) of our results, we use the deviation between $\rho^\text{exp}$ and $\rho^\text{teo}$, under a proper modulus definition for matrices. Then, 
\begin{equation}
	\delta(\%) = \frac{||\rho^\text{teo}-\rho^\text{exp}||}{||\rho^\text{teo}||}\times 100(\%),
\end{equation}
where for $||A||$ we use
\begin{equation}
	||A|| = max(a_1,..,a_n),
\end{equation}
and $\{a_1,..,a_n\}$ are the singular values of $A$.\par 
The results for the three experiments are given below, as well as the theoretical form of the deviation matrix, $\rho^\text{teo}$, and the experimental errors, $\delta$.
\begin{itemize}
	\item \textit{Pauli-Z} gate ($S_1$)
\end{itemize}
\begin{equation}
	\rho^\text{teo}_{S_1}=\mqty(1 & 0 \\ 0 & -1),
\end{equation}
\begin{align}
	\rho^\text{exp}_{S_1}&=  \mathcal{T}[S_1\rho S_1^\dag,\Pi_1S_1\rho S_1^\dag \Pi_1^\dag] \nonumber \\
	&= \mqty(0.9603 & -0.0501-i0.0822 \\ -0.0501+i0.0822 & -0.9603),
\end{align}
\begin{equation}
	\delta_1 \approx 5.02\%.
\end{equation}

\begin{itemize}
	\item \textit{Pauli-X} gate ($S_2$)
\end{itemize}
\begin{equation}
	\rho^\text{teo}_{S_2}=\mqty(-1 & 0 \\ 0 & 1),
\end{equation}
\begin{align}
	\rho^\text{exp}_{S_2}&= \mathcal{T}[S_2\rho S_2^\dag,\Pi_1S_2\rho S_2^\dag \Pi_1^\dag] \nonumber \\
	&=\mqty(-0.9342 & 0.0059-i0.0007 \\ 0.0059+i0.0007 & 0.9342),
\end{align}
\begin{equation}
	\delta_2 \approx 7.71\%.
\end{equation}

\begin{itemize}
	\item \textit{Hadamard} gate ($S_3$)
\end{itemize}
\begin{equation}
	\rho^\text{teo}_{S_3}=\mqty(0 & 1 \\ 1 & 0),
\end{equation}

\begin{align}
	\rho^\text{exp}_{S_3} &=\mathcal{T}[S_3\rho S_3^\dag,\Pi_1S_3\rho S_3^\dag \Pi_1^\dag] \nonumber \\
	&=\mqty(-0.1104 & 1.0221-i0.1443 \\ 1.0221+i0.1443 & 0.1104),
\end{align}

\begin{equation}
	\delta_3 \approx 10.79\%.
\end{equation}

The results are accurate, as can be noted from the expected form of the matrices above. The deviation matrix positions where a "1" is expected, have a much bigger element than all the others, as well as a small (nearly zero) imaginary part, whereas the others should have zero real and imaginary parts. These approach zero, when compared to the other non-zero elements, but the small deviations are totally expected. We predict a large number of potential errors, such as imperfections and inhomogeneities in the static and RF magnetic fields, and pulse phase calibration errors, which will be discussed more carefully in the end. Note that the tomography procedure increases the number of operations, and thus some possible inaccuracies might come along.
\subsection{Two-qubit system}
For the second case, a NMR routine, proposed by Knill et al. was followed \cite{knill}. It consists of a sequence of pulses, known as \textit{temporal averaging}. In this case, the room temperature density matrix for the composite spin system of the hydrogen (proton) and carbon-13 nucleus is approximately given by
\begin{widetext}
	\begin{align}
\rho_0 = &\left(
\begin{array}{cccc}
1 & 0 & 0 & 0\\
0 & 1 & 0 & 0 \\
0 & 0 & 1 & 0\\
0 & 0 & 0 & 1\\
\end{array}
\right) + \frac{\hbar B_0}{2k_B T}\left(
\begin{array}{cccc}
\gamma^1+\gamma^2 & 0 & 0 & 0\\
0 & \gamma^1-\gamma^2 & 0 & 0 \\
0 & 0 & \gamma^2-\gamma^1 & 0\\
0 & 0 & 0 & -\gamma^1-\gamma^2\\
\end{array}
\right), \nonumber \\  
\approx & \left(
\begin{array}{cccc}
1 & 0 & 0 & 0\\
0 & 1 & 0 & 0 \\
0 & 0 & 1 & 0\\
0 & 0 & 0 & 1\\
\end{array}
\right) + 5.86\times 10^{-5}\left(
\begin{array}{cccc}
1 & 0 & 0 & 0\\
0 & 0.5981 & 0 & 0 \\
0 & 0 & -0.5981 & 0\\
0 & 0 & 0 & -1\\
\end{array}
\right),\label{dev} 
	\end{align}

\end{widetext}
where $\gamma^1\approx42.57 \ \text{MHz T}^{-1}$ and $\gamma^2 \approx  10.71  \ \text{MHz T}^{-1}$ are the gyromagnetic ratios of the hydrogen and carbon-13 nucleus, respectively, and $T=20^\circ$C. Eq.\eqref{dev} clearly doesn't correspond to a pseudo-pure state. Once all NMR observables arise from the deviation matrix, we focus our attention of the second term of Eq.\eqref{dev}, since the first one will drop out of our calculations. 
We are interested in preparing a sequence of pulses that renders a pseudo-pure state after its application. Unfortunately, this is not possible to achieve with an isolated sequence, which means that no single sequence $P_i$ can produce a pseudo-pure state, $P_i \rho_0 P_i^\dag$, for the equilibrium density matrix of Eq.\eqref{dev}. Therefore, we perform three sequences, in the form of theree operators $P_0$, $P_1$ and $P_2$. If we choose carefully these sequences, the sum of the end results,
\begin{equation}
	\rho_{\text{pps}} = \frac{1}{3}\big(P_0 \rho_0 P_0^\dag + P_1 \rho_0 P_1^\dag + P_2 \rho_0 P_2^\dag\big) ,\label{cr}
\end{equation}
 is a pseudo-pure state (pps). These matrices are 
 \begin{equation}
	P_0 = \left(
\begin{array}{cccc}
1 & 0 & 0 & 0\\
0 & 1 & 0 & 0 \\
0 & 0 & 1 & 0\\
0 & 0 & 0 & 1\\
\end{array}
\right),
\end{equation}

 \begin{equation}
	P_1 = \left(
\begin{array}{cccc}
 0 & -i & 0 & 0 \\
 0 & 0 & -1 & 0 \\
 -1 & 0 & 0 & 0 \\
 0 & 0 & 0 & i \\
\end{array}
\right),
\end{equation}

 \begin{equation}
	P_2 =\left(
\begin{array}{cccc}
 0 & 0 & -1 & 0 \\
 -i & 0 & 0 & 0 \\
 0 & -1 & 0 & 0 \\
 0 & 0 & 0 & i \\
\end{array}
\right),
	\end{equation}
The first matrix $P_0$ is the identity matrix, such that no sequence must be applied. Using the equilibrium density matrix of \eqref{dev} into Eq.\eqref{cr}, it gives
\begin{equation}
	\rho_\text{pps} = \xi_1\left(
\begin{array}{cccc}
1 & 0 & 0 & 0\\
0 & 1 & 0 & 0 \\
0 & 0 & 1 & 0\\
0 & 0 & 0 & 1\\
\end{array}
\right) + \xi_2\left(
\begin{array}{cccc}
0 & 0 & 0 & 0\\
0 & 0 & 0 & 0 \\
0 & 0 & 0 & 0\\
0 & 0 & 0 & 1\\
\end{array}
\right), \label{dev2} 
\end{equation} 

where $\xi_1 \approx  1 + 1.95\times 10^{-5} $ and $\xi_2=-7.81\times 10^{-5}$. Henceforth, we are able to perform computations on the $\ket{11}$ state, without undesirable contaminations of other states, since the first term on the RHS of Eq.\eqref{dev2} adds no contribution to the magnetization. Having the explicit form of the matrices needed to create a pseudo-pure state, it remains the problem of experimental realising those matrices onto physical pulse sequences. The solution to this is also described in Ref. \cite{knill}. During the time interval of each pulse, the total Hamiltonian that corresponds to the time evolution of the system is that of Eq.\eqref{2spinH}. Nevertheless, since $J\approx208$Hz, then the relation $\omega_1^1,\omega_1^2 \gg 2\pi J$ holds, and we can discard the $J-$coupling term, as mentioned before, during the pulse applications. The Hamiltonian reduces to 
\begin{align}
	\mathcal{H}_{\text{rot}}&[\omega_1^1,\phi^1,\omega_1^2,\phi^2] = - \hbar\omega_1^1\big(\cos\phi^1  \ I_x^1\otimes\mathbb{1}^2  \nonumber\\
	&- \sin\phi^1 \ I_y^1\otimes\mathbb{1}^2\big)-  \hbar\omega_1^2\big(\cos\phi^2 \  \mathbb{1}^1\otimes I_x^2\nonumber \\
	& - \sin\phi^2 \  \mathbb{1}^1\otimes I_y^2 \big) .
\end{align}
For the time intervals between pulses, the Hamiltonian simply includes the J-coupling term,
\begin{equation}
\mathcal{H}_{\text{rot}} =	\hbar 2\pi J \big(I_z^1\otimes I_z^2\big).
\end{equation}
The total pulse sequences $P_1$ and $P_2$ can, then, be written as \begin{align}
	P_1 &=  e^{-i\Delta t_2\omega_1^2\mathbb{1}^1\otimes I_y^2} \  e^{-i\Delta t 2\pi J (I_z^1\otimes I_z^2)} \nonumber\\
	&\times e^{-i\Delta t_2\omega_1^2\mathbb{1}^1\otimes I_x^2} \  e^{-i\Delta t_1\omega_1^1 I_y^1\otimes\mathbb{1}^2} \nonumber \\
	&\times e^{-i\Delta t 2\pi J (I_z^1\otimes I_z^2)} \  e^{-i\Delta t_1\omega_1^1 I_x^1\otimes\mathbb{1}^2}, \\  
	P_2 &= e^{-i\Delta t_1\omega_1^1 I_y^1\otimes\mathbb{1}^2} \  e^{-i\Delta t 2\pi J (I_z^1\otimes I_z^2)} \nonumber\\
	&\times e^{-i\Delta t_1\omega_1^1 I_x^1\otimes\mathbb{1}^2} \  e^{-i\Delta t_2\omega_1^2\mathbb{1}^1\otimes I_y^2} \nonumber \\
	&\times e^{-i\Delta t 2\pi J (I_z^1\otimes I_z^2)} \  e^{-i\Delta t_2\omega_1^2\mathbb{1}^1\otimes I_x^2}  , 
\end{align}
where the time intervals $\Delta t$, $\Delta t_1$ and $\Delta t_2$ are determined by the relations 
\begin{align}
	\Delta t  \ J = \pi/2, \label{time1}\\
	\Delta t_1 \ \omega_1^1 = \pi/2 , \label{time2} \\
	\Delta t_2 \ \omega_1^2 = \pi/2 \label{time3}.
\end{align}
This experimental method, although solves the problem, brings additional complications to the process, because from now on, all the computations must be applied, independently, on three experiments, and those results must be added in the end, to mimic its application on a single state $\ket{11}$. This triples the experimental effort, in the case of a two-spin system. We expect that the potential errors will also be increased. This process is not unique, and other methods could be used. However, they would require experimental tools that we didn't have at our disposal, such as field gradient techniques. See Ref. \cite{pps2} for an alternative solution to this problem. \par
Now, we present the results for the preparation of the pseudo-pure state, following this method. The tomography function has now a more complex argument, namely, $\mathcal{T}[\rho,\Omega_1\rho\Omega_1^\dag,\Omega_2\rho\Omega_2^\dag,\Omega_3\rho\Omega_3^\dag,\Omega_4\rho\Omega_4^\dag]$, and must be applied after each change in the density matrix $\rho$. The notation follows from previous chapters. To compare the expected with the experimental results, we need to rewrite Eq.\eqref{dev2} as
\begin{equation}
	\rho_\text{pps} = \xi_1' \mathbb{1} + \xi_2'\  diag(1/3,1/3,1/3,-1),
\end{equation} 
where $\xi_1'=1$ and $\xi_2'=5.86\times 10^{-4}$, and note that the first term is the identity matrix, which won't contribute to the detected signal, and can be discarded. Due to the normalization of the data, we should compare our experimental result with the following theoretical form
\begin{equation}
	\rho_\text{pps}^\text{teo} =  \left(
\begin{array}{cccc}
 0.3333 & 0 & 0 & 0\\
 0 & 0.3333 & 0 & 0 \\
 0 & 0 & 0.3333 & 0\\
 0 & 0 & 0 & -1\\
 \end{array}
\right).
\end{equation} 
The experimental result $\rho_\text{pps}^\text{exp}$ is composed by the sum of the three experiments, corresponding to the applications of $P_0$, $P_1$ and $P_2$,
\begin{widetext}
\begin{align}
\rho_\text{pps}^\text{exp}&= \frac{1}{3}\big(\mathcal{T}[\rho,\Omega_1\rho\Omega_1^\dag,\Omega_2\rho\Omega_2^\dag,\Omega_3\rho\Omega_3^\dag,\Omega_4\rho\Omega_4^\dag] \  + \  \mathcal{T}[P_1\rho P_1^\dag ,\Omega_1P_1\rho P_1^\dag \Omega_1^\dag,\Omega_2 P_1^\dag\rho P_1^\dag\Omega_2^\dag,\Omega_3 P_1^\dag\rho P_1^\dag \Omega_3^\dag,\Omega_4 P_1\rho P_1^\dag\Omega_4^\dag] \nonumber \\
&+ \mathcal{T}[P_2\rho P_2^\dag,\Omega_1P_2\rho P_2^\dag\Omega_1^\dag,\Omega_2P_2\rho P_2^\dag\Omega_2^\dag,\Omega_3P_2\rho P_2^\dag\Omega_3^\dag,\Omega_4P_2\rho P_2^\dag\Omega_4^\dag]\big), \nonumber \\
&=\left(
\begin{array}{cccc}
 0.2844-0.0253 i & -0.05497-0.09281 i & 0.03786-0.02695 i & 0.01413+0.03190 i \\
 -0.05497+0.09281 i & 0.3408+0.0115 i & -0.01169+0.01876 i & -0.07782-0.01776 i \\
 0.03786+0.02695 i & -0.01169-0.01876 i & 0.2990+0.0013 i & -0.0383-0.2076 i \\
 0.01413-0.03190 i & -0.07782+0.01776 i & -0.0383+0.2076 i & -0.9242-0.0615 i \\
\end{array}
\right).
\end{align}
\end{widetext}
At this stage, the experimental error, using previous definitions, gives
\begin{equation}
	\delta_\text{pps} \approx 23.41\%.
\end{equation}
This value is large, but can be attributed to measurement imperfections, which are discussed later. We can also reconstruct the NMR wave-function, from $\rho_\text{pps}^\text{exp}$, as described in Sec. \ref{densitymatrix}. We expect the state
\begin{align}
\ket{\psi}_\text{pps}^\text{teo} \sim \ket{11} , \label{psiteo}
\end{align}
and, experimentally, we obtained
\begin{align}
\ket{\psi}_\text{pps}^\text{exp} = &0.0030 \times e^{i1.15}\ket{00} + 0.0001\times e^{-i2.91}\ket{01}\nonumber \\
&+0.0011\times e^{-i1.75} \ket{10} + 1.5853 \ket{11} .\label{psiexp}
\end{align}
The coefficient of the $\ket{11}$ state is noticeably larger than all the others, thus indicating the correction of the experimental preparation. Note that both the results of Eq.\eqref{psiteo} and \eqref{psiexp} are called NMR wave-functions, once a proper identity matrix multiplied by a factor was discounted in each of the density matrices $\rho_\text{pps}^\text{teo}$ and $\rho_\text{pps}^\text{exp}$, which is not measured by the NMR signal. This means that these states are artificial states, because a background identity matrix is subtracted from them. However, from the NMR point of view, these are the effective states we have available.
\par  
Then, we applied the \textit{Hadamard} quantum gate, $H=S_3\otimes\mathbb{1}^2$, where $S_3$ is the one-qubit \textit{Hadamard} defined before, accomplished with the transformation  
\begin{equation}
	H = e^{-i\Delta t_2 \mathcal{H}_\text{rot}[\omega_1^1,0,0,0]/\hbar} \ e^{-i\Delta t_1 \mathcal{H}_\text{rot}[\omega_1^1,\pi,0,0]/\hbar},
\end{equation}
where the time intervals are determined by 
\begin{align}
	\Delta t_1 \omega_1^1 &= \pi/2,\\
	\Delta t_2 \omega_1^1 &= \pi.
\end{align}
In matrix notation, it gives
\begin{equation}
	H=\frac{1}{\sqrt{2}}\left(
\begin{array}{cccc}
 1 & 0 & 1 & 0 \\
 0 & 1 & 0 & 1 \\
 1 & 0 & -1 & 0 \\
 0 & 1 & 0 & -1 \\
\end{array}
\right).
\end{equation}
Therefore, $H$ acts on the hydrogen spin, and transforms the basis states $\ket{0\sigma}$ and $\ket{1\sigma}$ as  
\begin{align}
	\ket{0\sigma} &\rightarrow \frac{\ket{0\sigma}+\ket{1\sigma}}{\sqrt{2}} ,\\
	\ket{1\sigma} &\rightarrow \frac{\ket{0\sigma}-\ket{1\sigma}}{\sqrt{2}},
\end{align}
where $\sigma$ is any state of the carbon-13 spin. The theoretical density matrix is given by
\begin{align}
	\rho_{\text{pps},H}^\text{teo} &= H\rho_\text{pps}^\text{teo}H^\dag \nonumber\\
	&=\left(
\begin{array}{cccc}
 0.3333 & 0 & 0 & 0 \\
 0 & -0.3333 & 0 & 0.6667\\
0 & 0 & 0.3333 & 0\\
0& 0.6667 & 0 & -0.3333 \\
\end{array}
\right).\label{Hpps}
\end{align} 
For the experimental result, we obtained 
\begin{widetext}
	\begin{align}
	\rho_{\text{pps},H}^\text{exp} = & \frac{1}{3}\Big(\mathcal{T}[H\rho H^\dag,\Omega_1H\rho H^\dag\Omega_1^\dag,\Omega_2H\rho H^\dag\Omega_2^\dag,\Omega_3H \rho H^\dag\Omega_3^\dag,\Omega_4H\rho H^\dag\Omega_4^\dag] \nonumber \\
	  &+   \mathcal{T}[H P_1\rho  P_1^\dag H^\dag ,\Omega_1 H P_1\rho  P_1^\dag H^\dag \Omega_1^\dag,\Omega_2 H P_1^\dag\rho P_1^\dag H^\dag \Omega_2^\dag,\Omega_3 H P_1^\dag\rho P_1^\dag H^\dag \Omega_3^\dag,\Omega_4 H P_1\rho P_1^\dag H^\dag \Omega_4^\dag] \nonumber \\
&+ \mathcal{T}[HP_2\rho P_2^\dag H^\dag,\Omega_1HP_2\rho P_2^\dag H^\dag \Omega_1^\dag,\Omega_2 H P_2\rho P_2^\dag H^\dag \Omega_2^\dag,\Omega_3 H P_2\rho P_2^\dag H^\dag\Omega_3^\dag,\Omega_4H P_2\rho P_2^\dag H^\dag \Omega_4^\dag]\Big), \nonumber \\
	=&\left(
\begin{array}{cccc}
 0.2985-0.0527 i & -0.0179+0.1592 i & 0.03942+0.03982 i & -0.01039-0.07239 i \\
 -0.0179-0.1592 i & -0.3825-0.0360 i & -0.05330+0.11672 i & 0.4873-0.0725 i \\
 0.03942-0.03982 i & -0.05330-0.11672 i & 0.2856-0.0056 i & 0.0607+0.1693 i \\
 -0.01039+0.07239 i & 0.4873+0.0725 i & 0.0607-0.1693 i & -0.2016+0.0004 i \\
\end{array}
\right).
\end{align} 
\end{widetext}
As before, the result seems accurate, as it clearly shows a strong similarity with the $\rho_{\text{pps},H}^\text{teo}$. Nevertheless, some matrix elements show a few deviations. The experimental error for the \textit{Hadamard} quantum gate is 
\begin{equation}
	\delta_{\text{pps},H} \approx 33.78\%.\label{HHH}
\end{equation}
Note that, from the definition of Eq.\eqref{Hpps}, $\delta_{\text{pps},H}\approx \delta_{pps} +  2det(H)\delta_{H}$, $det$ is the determinant, and $\delta_H$ is the effective error for the experimental creation of $H$. The total error of Eq.\eqref{HHH} has a contribution from the previous step, so $\delta_{\text{pps},H} \geq \delta_\text{pps}$, whereas in the second term, although $det(H)=1$ theoretical, the experimental $H$ must be used, for which the determinant is not exactly one. The NMR wave function can also be calculated. The theoretical and experimental results are 
\begin{align}
	\ket{\psi}_\text{pps,H}^\text{teo} &= H \ket{\psi}_\text{pps}^\text{teo}\nonumber \\
	&\sim (\ket{01} - \ket{11}), \label{Hwf}
\end{align}
\begin{align}
	\ket{\psi}_\text{pps,H}^\text{exp} &= 0.0040 \times e^{i1.42} \ket{00} + 0.5137 \times e^{i 2.99} \ket{01} \nonumber \\ 
	&+ 0.0023 \times e^{-i1.92} \ket{10} + 0.2862 \ket{11}.
\end{align}
At this stage, both the amplitude and the phase of each coefficient are important, as we need to ensure that, as suggested by Eq.\eqref{Hwf}, the phase difference between the $\ket{01}$ and $\ket{11}$ coefficients is approximately $\pi$. Experimentally, $\theta_{\ket{01}} -\theta_{\ket{11}} \approx 2.99$, where $\theta_{\ket{a}} = Arg(\bra{a}\ket{\psi})$, which is close enough, given the experimental errors. Furthermore, the amplitudes of the expected states $\ket{01}$ and $\ket{11}$ are much larger than that of the two remaining states, as they should. 
\section{Discussion and conclusions}
As mentioned in previous sections, the results are very satisfactory, although small errors were encountered. Those deviations are expected, given the experimental conditions under which the experiments were conducted.\par 
 First of all, it is important to understand how pulse calibration errors are expected to affect the results, as well as how are the pulse time-lengths determined (i.e, how are equations like are experimentally Eq.\eqref{time1}-\eqref{time3} satisfied). To do so, a first experiment was performed, were a $\pi-$pulse was applied, in each channel. It is known that, if in an equilibrium situation, the observable transverse magnetization must vanish completely after $\pi-$pulse (in either the $x$ or $y$ direction), once the bulk magnetization is rotated from the $z$ direction, to the $-z$ direction. With this in mind, we span the pulse length in some interval, until we reach the pulse length $\Delta t$ that corresponds to zero transverse magnetization. We can further use that time interval to obtain any pulse length, because it scales linearly (e.g, a $\pi/2-$pulse corresponds to $\Delta t/2$). This procedure must be applied to both nucleus, because different gyromagnetic constants yield different Larmor frequencies, and then different time intervals. There is a potential for error accumulation at this phase, since a zero magnetization is never reach, in practise. A small deviation in the pulse length calibration compromises the entire experiment, because once calculated in the begining, the same time intervals are used throughout all NMR sequences. On the second hand, the signal from the carbon-13 molecules is $\sim 100$ times weaker, due to the relatively low percentage of its host molecule in the chloroform sample. This difference brings essentially two major problems, from the experimental point of view: first, the signal for the case of the two-spin experiments (where the carbon-13 molecules were used) is, itself, very reduced, and close, in some cases, to the thermal noise, which lowers the accuracy of the measurement; second, the signals of the one-spin and two-spin experiments overlap (see Fig.\ref{spectra}), such that the integrations of the peaks might be contaminated from one experiment to the other. Those two problems also explain the errors, and mostly, the accuracy difference between the first and second experiments. \par 
 It is also worth mention that, due to the lack of time available in the laboratory, no optimisation operations were implemented. These operations would have been necessary to effectively lower the experimental errors, once the required accuracy for quantum computing is way higher than that needed for the traditional NMR routines. These would include a considerable number of operations, such as field-homogeneity tests, to ensure that the condition $\bm{\nabla}\cdot \mathbf{B}_0 \approx \mathbf{0}$ holds (this happens to be essential for the validity of the calculations, and haven't been verified). Furthermore, a better calibration method, used for the pulse-length calculations, could have been implemented, once it is important for the Larmor frequencies to be calculated with an extremely high accuracy. We note that, above all these reasons, we still had to deal with the fact that the NMR signal is weak by definition, because the individual nuclear magnetic moments are very small, when compared with those of electrons, and the distribution of nuclear magnetic moments is nearly isotropic. This means that it will always be impossible to eradicate errors in such a sensitive experiment, as NMR seems to be. Besides, with the increase of the number of qubits, the number of operations increases exponentially, which bring additional experimental complications. \par 
 Nevertheless, we would like to mentioned that, although not perfectly accurate, this technique provides a very pedagogical and physical way to approach quantum computing. Within this approach, it is possible, with a few number of external parameters, to perform unitary transformations onto a reduced wave-function, which is something, in principle, not obvious of how to produce, essentially because quantum signals are, most of the times, weak. Although not the best solution to quantum communication, once NMR sequences are local experiments that are difficult to extend, the current method seems to be a very practical way to physically perform quantum algorithms, paving the way to the construction of better and better quantum computers. From the scalability point of view, the NMR approach is controversy, given that an increase in the number of qubits would require two things: a large amount of operations, such that a way of automation would be required, also because the experiments rapidly become very time consuming; a sample with molecules with N spins, to prepare a N-qubit system. Last requirement eventually implies large molecules, which must also remain independent, for liquid samples. For a number of, e.g, $N=100$ qubits, it is already difficult find a mixture composed by independent molecules with 100 active spins, and so scaling-up NMR capacities to produce devices capable of solving real problems has been  questioned. Asking if those problems are intrinsic to the number of qubits in the NMR quantum computer, or can be attributed to the NMR technology is still an open question nowadays \cite{sca1,sca2,sca3}.\par 
 To conclude, from the student's point of view, this work was really interesting and challenging, as it allowed for a distinct approach to previous acquired knowledge, and use it to perform experimental operations, which is not the student's usual experience. NMR quantum computing seems to be a very powerful way of constructing small and local devices, capable of operating at the quantum level, and can be utilised to perform quantum algorithms and solve small N-body problems. Although some limitations may come along, such as those identified before, concerning inherent problems of scalability, there is still plenty of room for improvements, from the technological point of view. With the progressive advance of the scientific knowledge that is predictable for the next years, NMR quantum computing might become a promising tool for future quantum-based processing devices. \vspace{-0.29cm}
 \section{Acknowledgments}
 This work was funded by the New Talents in Quantum Technologies 2020 Programme, delivered by the Calouste Gulbenkian foundation, to which we address our gratitude for all the support during the internship. The student also thanks his supervisor, Prof. Jo\~{a}o Lu\'{i}s Figueirinhas, for his essential help, guidance, and patience throughout the work.

\bibliographystyle{apsrev4-1}
\bibliography{report.bib}
\bibliographystyle{apsrev4-1}

\end{document}